\documentclass[aps,prb,floatfix,twocolumn]{revtex4}
\usepackage{graphicx}
\usepackage{amsmath}
\usepackage{amssymb}
\newcommand{\ti}{t_{\rm i}}
\newcommand{\tf}{t_{\rm f}}
\newcommand{\hH}{\hat H}
\newcommand{\Hi}{\hat H_{\rm i}}
\newcommand{\Hf}{\hat H_{\rm f}}
\newcommand{\Ri}{R_{\rm i}}
\newcommand{\Rf}{R_{\rm f}}

\newcommand{\BEQ}{\begin{eqnarray}}
\newcommand{\EEQ}{\end{eqnarray}}
\newcommand{\BEA}{\begin{eqnarray}}
\newcommand{\EEA}{\end{eqnarray}}
\renewcommand{\H}{{\hat H}}

\newcommand{\V}{\hat V}

\renewcommand{\d}{{\rm d}}

\newcommand{\ep}{\varepsilon}

\newcommand{\eps}{\varepsilon}

\newcommand{\tr}{{\rm tr}}

\newcommand{\W}{{\cal W}}

\newcommand{\p}{\partial}

\newcommand{\U}{U} 

\renewcommand{\S}{{S_{\rm th}}}

\newcommand{\half}{\frac{1}{2}}

\newcommand{\aad}{{\tilde a}}
\def\dbarrm {{\mathchar'26\mkern-11mu{\rm d}}}                       %
\begin{document} 
\draft
\title
{Quantum thermodynamics: thermodynamics at the nanoscale}
\author{A.E. Allahverdyan$^{1,2)}$, R. Balian$^{3)}$
 and Th.M. Nieuwenhuizen$^{1)}$}

\address{
$^{1)}$ Institute for Theoretical Physics,
Valckenierstraat 65, 1018 XE Amsterdam, The Netherlands\\
$^{2)}$Yerevan Physics Institute,
Alikhanian Brothers St. 2, Yerevan 375036, Armenia\\ 
$^{3)}$ SPhT, CEA-Saclay, 91191 Gif-sur-Yvette cedex, France}

\begin{abstract}
A short introduction on quantum thermodynamics is given and three new
topics are discussed: 1) Maximal work extraction from a finite quantum
system.  The thermodynamic prediction fails and a new, general result
is derived, the ``ergotropy''. 2) In work extraction from
two-temperature setups, the presence of correlations can push the
effective efficiency beyond the Carnot bound.  3) In the presence of
level crossing, non-slow changes may be more optimal than slow ones.
\end{abstract}


\maketitle

Thermodynamics originated in the nineteenth century as a science of
engines. It took a while before it was understood that its laws 
originate from a molecular picture and, with it, 
the theory at that scale, quantum mechanics. Still, these efforts 
dealt with macroscopic systems.

When going to the limits of thermodynamics, one may wonder whether it
exists for finite systems in the quantum regime, like few level atoms.
At first sight, this seems a hopeless task. After all, there is
nothing thermodynamical in the groundstate of the hydrogen atom, so
why should there be such behavior for other finite systems?  But let
us notice that a thermodynamic setup deals with three parts: the
system under consideration, a bath and a work source.  The latter both
have to remain extensive for having the proper physical meaning: the
work source transfers high-graded energy (energy-without-entropy),
while the bath transfers low-graded energy related to uncontrollable,
thermalized degrees of freedom.  In this setup there is a hope to
understand thermodynamical laws directly from quantum mechanics.

One of 
the first question is then: can the energy $U$ of the system S be 
uniquely identified, even though there is a non-small coupling to the bath?
We have shown before that this can be done in two well known
system+bath models: the Caldeira-Leggett model of a damped quantum
particle in an external potential and the spin-boson model of a spin
in the presence of a bath. 
For the Caldeira-Leggett model the Hamiltonian reads
\BEA
\label{hamiltonian}
\H_{\rm tot}&=&\H+\H_B+\H_I,\quad 
\H=\frac{\hat p^2}{2m}+V(\hat x)\\
\H_B&=&\sum_{i}\left [
\frac{\hat p_i^2}{2m_i}+\frac{m_i\omega_i^2}{2}\hat x_i^2\right],
\quad \H_I=-\sum_i c_i \hat x_i \hat x
\nonumber
\EEA
and we choose a dense quasi-Ohmic spectrum
\BEQ J(\omega)=\frac{\pi}{2}\sum_i \frac{c_i^2}{m_i\omega_i}
\delta(\omega-\omega_i)=
\frac{\gamma \omega\Gamma^2}{\omega^2+\Gamma^2} \EEQ
where $\gamma$ is the coupling or damping constant and
$\Gamma$ the `Debye' cutoff frequency.
Studying the Langevin dynamics, it was seen that for times larger
than the inverse Debye frequency $1/\Gamma$, the unitary part of the
equations of
motion is given by the Hamiltonian~\cite{Weiss,NAlinw}
\BEQ 
\H_S=\frac{\hat p^2}{2m}+\V(\hat x)-\half \gamma\Gamma \hat x^2.
\EEQ
For the harmonic case $\V=\half b\hat x^2$ this implies an 
effective spring constant $a=b-\gamma\Gamma$ 
and $b>\gamma\Gamma$ or $a>0$.
This is exactly what is needed for stability of the overall system.

Now the internal energy is {\it uniquely} defined as $U=\tr \hat\rho
\H_S$, where $\hat\rho$ is the density matrix. A similar approach
works~\cite{ANspinboson} in the spin-boson model ~\cite{Weiss}, the
most important difference being that the (bare) Hamiltonian of the
system already does the job.

The action of a {\it macroscopic work} source on the system S will be
to make a parameter (for instance the effective mass $m$ or the
spring constant $b$) time-dependent. Therefore
also the work $\dbarrm W$ done on S can be uniquely defined,
as $\dbarrm W=\tr\hat\rho \,
\d\H=\tr\hat\rho\, [(\p\H/\p m)\d m+(\p\H/\p b)\d b]$.
It coincides with minus the energy change of the work-source.

Having these two ingredients, the first law is established,
with the heat added to S: $\dbarrm Q=\d U-\dbarrm W$.
This indicates that {\it Quantum Thermodynamics
exists as a non-trivial subject}.

Now we can go to the second law. In the absence of a thermodynamic
limit, its many formulations have different domains of validity.
So one can no longer speak of {\it the} second law and, moreover,
{\it entropy is a bad indicator} for a variety of  reasons, see below.

In the PQE-2003 proceedings it was pointed out that~\cite{PQE03}:

a). A generalized Thomson formulation (cycles cost work) is 
generally valid, provided the
total system (including the bath and coupling to it)
starts in gibbsian equilibrium~\cite{ANthomson}.
An example of this case is to make a cyclic change in $b$. ~\cite{NAlinw}

b). The Clausius inequality $\dbarrm Q\le T\d S$ can be violated.
This violation sets in at high $T$ in subdominant $\hbar^2/T$ terms,
and becomes strongest at $T=0$, where $T\d S$ vanishes.
The physical mechanism is the formation of a cloud of bath modes
around the central system, as occurs for polarons and Kondo
systems. Energy of this cloud can enter $\d U$ as heat. 
Experimental tests have been designed for quantum electronics
~\cite{ANPRB} and quantum optics~\cite{ANspinboson}.

c). The rate of irreversible (non-adiabatic) work -- sometimes called
energy dispersion, when referring to the energy of the work source --
may be negative at low enough $T$.  This contrasts the classical case
where it is equal to $T$ times the rate of entropy production.  As a
result, starting out of equilibrium, a finite number of cycles could
be designed where work is extracted from the bath
~\cite{ANspinboson,NAlinw}.

1). The first new problem we treat here is one of the most known ones in
thermodynamics: what is the maximal amount work that can be extracted from a
non-equilibrium finite quantum system via 
cyclic processes~\cite{ABNergotropy}? 

{\it The maximal work-extraction problem} is posed in the following
way \cite{landau,balian}.  Let a system S be in an initial 
state with density matrix $\hat\rho_0$ and have Hamiltonian $\hH$. Certain
external fields are exerted on S in the time-interval $[0,\tau]$,
which amount to a cyclic variation of the Hamiltonian: $\hH(0)
=\hH(\tau)=\hH$. As S is assumed to be thermally isolated (it moves under
external fields and its own dynamics), the work $W$ done by external
sources is \cite{landau,balian}
\BEA
\label{s1}
W={\rm tr}[\, \hH\,\hat\rho_\tau \,]
-{\rm tr}[\, \hH\,\hat\rho_0  \,],
\EEA
where $\hat\rho_\tau$ is the state of S at the final moment $\tau$. Work
is extracted if $W<0$, and the questions is what is the maximal amount
$|W|$ which can be extracted, given $\hat\rho_0$ and $H$~\cite{landau,balian}.
The standard answer \cite{landau,balian} starts with postulating that 
the final state is in equilibrium at some 
temperature $T_{\rm f}=1/\beta_{\rm f}$,
\BEA
\label{odu}
\hat\rho_\tau=\hat\rho_{\rm f}
=\frac{e^{-\beta_{\rm f} \hH}}{Z}, 
\qquad Z={\rm tr}\,e^{-\beta_{\rm f}\hH}.
\EEA
The work is then $W_{\rm th}=U_{\rm f}(S_{\rm f})-U_{\rm
i}$, where $U_{\rm f}={\rm tr}\,[\hat\rho_{\rm f}\hH]$ is the final
(average) energy as a function of final entropy $S_{\rm f}=-{\rm
tr}\,\hat\rho_{\rm f}\ln \hat\rho_{\rm f}$, and $U_{\rm i}$ is the initial
energy. Since in the final equilibrium state
${\d U_{\rm f}}/{\d S_{\rm f}}=T_{\rm f}\geq 0$,
for $W_{\rm th}$ to be minimal, one should keep $S_{\rm f}$ as small
as possible.  As the entropy of a thermally isolated system can only
increase \cite{landau,balian}, the maximal amount of extracted work is
achieved for conserved entropy: 
$S_{\rm f}=S(0)=-{\rm tr}\,\hat\rho_0\ln \hat\rho_0$.  This
condition serves to determine $T_{\rm f}$.  Some known results follow:
{\it i)} When comparing two different initial
states $\hat\rho_0$ and $\hat\sigma_0$ of S having the same initial energy,
more work can be extracted from that one which has lower entropy.
{\it ii)} Conversely: if the maximal work extracted
from $\hat\rho_0$ is larger than that from $\hat\sigma_0$, 
$\hat\rho_0$ had a lower entropy. 
{\it iii)} If there is an independent, uncorrelated  system 
$\Omega$ (``spectator'')
with initial state $\hat\omega_0$ and Hamiltonian $\hH_{\omega}$, then it
just follows from the additivity of the entropy:
$S(\hat\rho_0\otimes\hat\omega_0)=
S(\hat\rho_0)+S(\hat\omega_0)$, that more work can be extracted from
$\hat\rho_0\otimes\hat\omega_0$ than from $\sigma_0\otimes\hat\omega_0$.

{\it Maximal work extraction in quantum mechanics}.
It is now our purpose to solve the maximal work extraction problem directly
from quantum mechanics, without involving any postulate.
Let us start from the spectral resolutions
\BEA
\label{gajl}
&&\hat{\rho}_0=\sum_{j\geq 1} p_j|p_j\rangle\langle p_j|,\qquad
\hat{H}=\sum_{k\geq 1} \eps_k|\ep_k\rangle\langle \ep_k|,
\EEA
where $\ep_k$, $p_k$, $|\ep_k\rangle$, $|p_k\rangle$ are 
the eigenvalues and orthonormal eigenvectors of $\hat\rho_0$ and $H$,
respectively.
We order the eigenvalues as
\BEA
\label{777}
p_1\geq p_2\geq\cdots,\qquad
\ep_1\leq\ep_2\leq\cdots 
\EEA
Since S is assumed to be thermally isolated, its evolution is
unitary: $\hat\rho_\tau=\hat U(\tau)\,\hat\rho_0\,\hat U^\dagger(\tau)$,
with $i\hbar\,{\d \hat{U}(t)}/{\d t}=\hH(t)\,\hat U(t)$.

The maximal extracted work $\W\equiv -W_{\rm max}$ was determined in
~\cite{ABNergotropy}:
\BEA
\label{ergotropy}
\! \W=U_0-\sum_{k\geq 1}p_k\,\ep_k, 
\quad U_0=\sum_{i,k\geq 1}|\,\langle p_k|\ep_i\rangle\,|^2\,\ep_i p_k.
\label{ru}
\EEA
For $\W$, which depends only on the initial state and Hamiltonian, we
coin the name {\it ergotropy}, from the ancient Greek, 
$\ep\rho\gamma o\tau\!\rho o\pi\eps\iota\alpha$ = work-transformation, 
transformation into work,  
($\ep\rho\gamma o\nu$, work; $\tau\!\rho o\pi\eta$, transformation, turn);
in analogy to Clausius' {\it entropy}, 
($\ep\nu- \tau\!\rho o\pi\eps\iota\alpha$, in-transformation).

It can be shown~\cite{ABNergotropy} that $\W_{\rm th}\ge\W\ge 0$, and
it is typically {\it not} equal to the thermodynamical expression
$\W_{\rm th}=U_{\rm i}-U_{\rm f}(S_{\rm f})$, because a unitary
dynamics conserves all eigenvalues of $\hat\rho_0$ and not only its
entropy. 
These additional conserved quantities are important for finite
systems, {\it the thermodynamically optimal state is typically
unreachable quantum mechanically}.  For macroscopic systems the effect
disappears or at least is expected to be very small -- in the same way
as energy is the only relevant constant of motion for a closed
macroscopic system \cite{landau,balian}.  It thus holds that $\W\leq
\W_{\rm th}$~\cite{ABNergotropy}.

The behavior of $\W$ can be rather different from that of its 
(unreachable) upper
bound $\W_{\rm th}$: {\it i)} states with a larger von Neumann entropy
may produce more work; {\it ii)} introduction of a spectator system can
invert the work-producing ability, that is, 
$\W(\hat\sigma)>\W(\hat\rho)$ can be consistent with
$\W(\hat\omega\otimes\hat\sigma)<\W(\hat\omega\otimes\hat\rho)$.
These and other related effects are discussed in detail in 
~\cite{ABNergotropy}.
The standard thermodynamic behavior is still valid for finite systems,
but only within the set of states that {\it majorize} one-another
~\cite{ABNergotropy}. In that case the standard von Neumann is the
proper indicator of work-providing ability.

2). {\it Work-extraction from a correlated two-temperature system
may exceed the Carnot bound.} 
The most traditional object of work-extraction is a pair of 
equilibrium systems
${\rm S_1}$ and ${\rm S_2}$ with Hamiltonians $\hH_1$ and $\hH_2$,
respectively.  The total Hamiltonian is $\hH=\hH_1+ \hH_2$ as ${\rm S_1}$
and ${\rm S_2}$ do not interact initially.
Initially, at $t=0$, ${\rm S_1}$ and ${\rm S_2}$ are
in equilibrium states 
\BEA
\label{w1}
\hat\rho_j(0)=\frac{e^{-\beta_j\hH_j}}{Z_j},
\qquad Z_j={\rm tr}\,e^{-\beta_j\hH_j},\qquad
j=1,2.
\EEA
at temperatures $T_2\geq T_1$. The total state of the baths
$\hat\rho_0$ may be correlated: though
$\hat\rho_{1,2}=\tr_{2,1}\hat\rho_0$, it may hold that
$\hat\rho_0\neq \hat\rho_1\otimes\hat\rho_2$.
The maximal work $|W|$ is extracted from $0$ till $\tau$, 
while for $t>\tau$ one switches on for the system $1$ ($2$) a weak interaction
with a thermal bath at temperature $T_1$ ($T_2$).  Due to this, both
systems separately relax back to their original states $\hat\rho_1(0)$ and
$\hat\rho_2(0)$ given by (\ref{w1}), though the total state need not return
to $\hat\rho_0$. 
Changes of energies during this relaxation
are completely attributed to heat. 
The heat $Q_1$ ($Q_2$) given by the first
(second) bath is:
\BEA
\label{heat}
Q_j={\rm tr}[\hH_j\,\hat\rho_j(0)]-{\rm tr}
[\hH_j\,\hat\rho_j(\tau)],\qquad j=1,2.
\EEA
This is equal to the final energy of the bath
minus the its initial value: the extracted work
comes from the baths.

We now present a new, general theorem for this setup. 
For any system S with Hamiltonian $\hH$ in a state
$\hat\rho$ with energy $\U ={\rm tr}[\hat\rho\,\hH]$,
one defines the thermodynamic entropy $ \S\equiv \beta U-\ln Z$
as the von Neumann entropy of an equilibrium state
$\hat\rho_{\rm eq}=\exp[-\beta \hH]/Z$ with the same energy. 
The temperature $T=1/\beta$  follows from 
${\d \S}/\d \U=\beta$.
Since ${\d^2 \S(\U)}/{\d \U ^2}=
{-1}/{\langle\,( \hH-\U\,)^2\,\rangle}<0$,
$\S (\U)$ is a monotonic and concave function.
Now the identity
$S_j(0)-S_j(\tau)=\S (U_j(0))- \S (U_j(\tau))$ for $j=1,2$ implies, 
using Eq. (\ref{heat}) and 
that $f(x)- f(y)\geq f'(x)(x-y)$ for any concave function $f$,
\BEA \label{burbon}
\S (U_j(0))- \S (U_j(\tau))
\geq\beta_j\,(\,U_j(0)-U_j(\tau)\,)=\beta_j Q_j.
\nonumber\\
\EEA
Here the $S_j=-{\rm tr}\,\hat\rho_j\ln\hat\rho_j$ are partial entropies,
while $S=-{\rm tr}\,\hat\rho_0\ln\hat\rho_0$ is the total entropy.  
Together they define the non-negative correlation entropy~\cite{balian}
\BEA \label{Scorr}
S_{\rm corr}=S_1+S_2-S\geq 0.
\EEA
Considering this relation at $t=0$ and $t=\tau$,
using $S(\tau)=S(0)$ (unitarity), eliminating
$Q_1=|W|-Q_2$ (energy conservation), and noting $Q_2\geq 0$ due to
$T_2\geq T_1$, 
one gets from (\ref{burbon}) as efficiency of the cycle
\BEA
\label{valois}\eta=\frac{ |W|}{Q_2}\le 		
(1-\frac{T_1}{T_2})+\frac{T_1}{Q_2}(S_{\rm corr,i}-S_{\rm corr,f}).
\EEA
If the initial state was factorized: $\hat\rho_0=\hat\rho_1(0)\otimes
\hat\rho_2(0)$, the maximal work extraction procedure will leave the final
state in a factorized form, so $S_{\rm corr,i}=S_{\rm corr,f}=0$, implying
the maximal possible efficiency is the Carnot bound. 
For a factorized $\hat\rho_0$, the total state of ${\rm S_1}$ and 
${\rm S_2}$ will return to it after interactions with the bath, so the
whole procedure can be repeated.
This is another formulation of the second law, but notice that our work 
extraction timescale, much smaller than the relaxation time of the bath,  
is not the most general one.

When the initial state is not factorized, the efficiency
(\ref{valois}) is not bound by Carnot: correlations can be traded for
efficiency.  Such a result was found before in a setup of a laser
cavity heated by a correlated three-level atom beam by Scully et
al.~\cite{Scully} and discussed at PQE 2003.

3). {\it Non-adiabatic variations may exceed the thermodynamic 
work performance}.
One of the formulations of the second law, the minimal work principle,  
states that work done on a thermally isolated equilibrium system is minimal 
for adiabatically slow (reversible) processes. 
Within the domain of finite quantum systems, it appears to be 
indeed generally valid provided the adiabatic energy levels do not cross. 
If level crossing does occur, this principle can be violated and
optimal processes are neither adiabatically slow nor reversible.
~\cite{ANlevelcrossing} 

{\it Formulation of the principle.}  
Consider a quantum system S which is thermally
isolated \cite{landau,balian,thirring}: it moves according to
its own dynamics and interacts with an external work source W.  This
interaction is realized via time-dependence of some parameters
$R(t)=[R_1(t),...,R_n(t)]$ of the system's Hamiltonian
$\hH(t)=\hH[R(t)]$. They move along a certain trajectory
$R(t)$ which at some initial time $\ti$ starts from $\Ri=R(\ti)$, and
ends at $\Rf$. The initial and final values of the Hamiltonian are
$\Hi=\hH[\Ri]$ and $\Hf=\hH[\Rf]$, respectively.  Initially S is assumed
to be in equilibrium at temperature $T=1/\beta$, that is, S is
described by a density operator:
\BEA
\label{gibbs}
\hat\rho(\ti)=\exp[-\beta \Hi]/Z_{\rm i},
\qquad
Z_{\rm i}={\rm tr}\,e^{-\beta \Hi}.
\EEA

One considers various processes which start from the same equilibrium state
(\ref{gibbs}) with $R$ moving between $\Ri$ and
$\Rf$ along a trajectory $R(t)$. 
The statement of the minimum work-principle then reads
\cite{landau,balian,thirring} 
\BEA
\label{2L}
W\geq W_{\rm ad} :
\EEA
The actual work $W$ done on S when moving along the trajectory $R(t)$
is not smaller than the work $W_{\rm ad}$ done during the adiabatically slow
variation between the same initial and final values $\Ri$ and $\Rf$
and along the same trajectory $R(t)$. By adiabatically slow we mean
a process which is realized with homogeneously vanishing velocity and
which thus takes very long time $\tf-\ti$, much longer than the
proper characteristic time of S.
For thermally isolated systems S adiabatically slow processes are
reversible. This is standard if S is macroscopic
\cite{landau,balian,thirring},
and was shown to be true also in the finite domain, where
the definition of reversibility extends unambiguously~\cite{ANlevelcrossing} 
(i.e., without using entropy).

Eq.~(\ref{2L}) is a statement on optimality: if the purpose of the
external source W is to extract work from S, then the actual work $W$
is negative, and to make it as negative as possible one proceeds with
very slow velocity.  On the other hand, if during some operation the
work has to be put into S, that is $W$ has to be positive, one tends
to minimize its amount, and again operates adiabatically slow.  In
macroscopic thermodynamics the minimum work principle is derived
\cite{landau,balian} from certain axioms which ensure that,
within the domain of their applicability, this principle is equivalent
to other formulations of the second law.  Derivations in the context
of statistical thermodynamics are presented in
\cite{jar,narnhofer}. 

{\it The minimal work principle for macroscopic systems} is proven in two
steps: first one establishes an inequality between the work and the
free energy difference, while in the second step one uses macroscopic
features of S to connect it with the adiabatic work.
One starts with the relative entropy 
${\rm tr}[\hat\rho(\tf)\ln\hat\rho(\tf)-
\hat\rho(\tf)\ln\hat\rho_{\rm eq}(\Hf)]$ between the
final state $\hat\rho(\tf)$ and an equilibrium
state $\hat\rho_{\rm eq}(\Hf)=\exp[-\beta \Hf]/Z_{\rm f}$, $Z_{\rm f}={\rm
tr}\,e^{-\beta \Hf}$. It has the same temperature $T=1/\beta$ as the
initial state $\hat\rho(\ti)$, but corresponds to the final Hamiltonian
$\Hf$. As follows from unitarity, 
${\rm tr}[\hat\rho(\tf)\ln\hat\rho(\tf)]= 
{\rm tr}[\hat\rho(\ti)\ln\hat\rho(\ti)]$. 
Combined with (\ref{gibbs}, \ref{s1}) and non-negativity of relative
entropy~\cite{balian}, this yields:
\BEA
\label{workfree-energy}
W\geq F(\Hf)-F(\Hi)\equiv T\ln {\rm tr}[e^{-\beta \Hi}]-
T\ln {\rm tr}[e^{-\beta \Hf}]), \nonumber
\EEA
where $F(\Hi)$ and $F(\Hf)$ are the free energies corresponding to
$\hat\rho(\ti)$ and $\hat\rho_{\rm eq}(\Hf)$, respectively.
There are several classes of macroscopic systems for which one can show
that the free energy difference $F(\Hf)-F(\Hi)$
indeed coincides with the adiabatic work \cite{narnhofer,jar,NAlinw}.

{\it Finite systems.}~\cite{ANlevelcrossing} 
For an arbitrary $N$-level quantum system S, 
Eq.~(\ref{workfree-energy}) does not have the physical meaning 
we need, since in general $F(\Hf)-F(\Hi)$ has no reason to coincide with
the the adiabatic work  needed in (\ref{2L}).
Therefore, one needs an independent,  possibly general, proof of 
(\ref{2L}). In ~\cite{ANlevelcrossing} we present one, 
starting from the quantum evolution equations
and using ideas from ~\cite{ANthomson}. 

Let the spectral resolutions of $\hH(t)$ and $\hat\rho(\ti)$ be
\BEA
\hat H(t)=\sum_{k=1}^N\ep_k(t)|k,t\rangle\langle k,t|,\quad
\langle k,t|n,t\rangle =\delta_{kn},\\
\label{khorezm}
\hat\rho(\ti)=\sum_{k=1}^N p_k|k,\ti\rangle\langle k,\ti|,\quad
p_k=\frac{e^{-\beta \ep_k(\ti)}}{\sum_n e^{-\beta \ep_n(\ti)}}.
\EEA
The ordering (\ref{777}) is satisfied at $t=\ti$.
For $\ti\leq t\leq \tf$ we expand on the complete set $|n,t\rangle$:
$\hat U(t)|k,\ti\rangle=\sum_{n=1}^N a_{kn}(t) \,
e^{-\frac{i}{\hbar}\int_{\ti}^{t}\d t'\,\ep_n(t')}\, 
|n,t\rangle$,
and uses (\ref{bek}, \ref{s1}) to get:
\BEA
\label{suomi1}
W-W_{\rm ad}=&&\sum_{m=1}^{N-1}[\ep_{m+1}(\tf)-\ep_{m}(\tf)]\,\Theta_m,\\
\Theta_m\equiv&&\sum_{n=1}^{m}\sum_{k=1}^N p_k
(\,|\aad_{kn}(\tf)|^2-|a_{kn}(\tf)|^2).
\label{suomi2}
\EEA
Assume now that the ordering (\ref{777}) is kept at $t=\tf$:
\BEA
\label{20'}
\ep_1(\tf)\leq ...\leq \ep_N(\tf).
\EEA
If energy levels did not cross each other, i.e. $\ep_{k+1}(t)-\ep_k(t)$
did not change its sign as a function of $t$ for
$\ti\leq t\leq \tf$, then (\ref{20'}) is implied by 
(\ref{777}).
According to non-crossing rule \cite{mead1}, if only one
independent parameter of the Hamiltonian $\hH(t)$ is varied, the above
condition is satisfied for any discrete-level quantum system:
level-crossing, even if it happens in model-dependent calculations
or due to approximate symmetry, does not survive arbitrary small
perturbation and is substituted by avoided crossing.  In that case the
standard adiabatic theorem of quantum mechanics
\cite{standardad} leads to
\BEA
\label{abu}
\aad_{kn}(\tf)=\delta_{kn}
\EEA
It follows from $|a_{kn}(\tf)|^2=|\langle n,\tf|\hat U|k,\ti\rangle|^2$
that
\BEA
\label{bek}
\sum_{k=1}^{N}|a_{kn}(\tf)|^2=\sum_{n=1}^{N}|a_{kn}(\tf)|^2=1.
\EEA
Combined with (\ref{777}, \ref{abu}) this brings $\Theta_m\ge 0$.
Together with Eqs.~(\ref{777}, \ref{20'}) this proves the 
minimum work principle (\ref{2L}) for finite systems,
without requiring a connection with the free energy 
difference~\cite{ANlevelcrossing}.

{\it Level crossing.} What happens if the adiabatic energy levels cross,
i.e. (\ref{20'}) is not valid? 
As an example we consider a spin-$1/2$ particle with Hamiltonian
~\cite{ANlevelcrossing}
\BEA
\label{sos}
\hH(t)=h_1(s)\hat\sigma_1-h_3(s)\hat\sigma_3,\qquad
s=t/\tau,
\EEA
where $\hat\sigma_1$ and $\hat\sigma_3$ are the corresponding Pauli matrices,
and $s$ is the reduced time with $\tau$ being the characteristic
time-scale.  The external magnetic fields $h_1$ and $h_3$ vary smoothly
in time.  Assume that {\it i)} for $s\to s_{\rm i}<0$ and for
$s\to s_{\rm f}>0$, $h_1(s)$ and $h_3(s)$ go to constant values
sufficiently fast; {\it ii)} for $s\to 0$ one has: $h_1(s)\simeq
\alpha_1 s^2$, $h_3(s)\simeq -\alpha_3 s$, where $\alpha_1$ and
$\alpha_3>0$ are constants.  {\it iii)} $h_1(s)$ and $h_3(s)$ are
non-zero for all $s$, $s_{\rm i}\leq s\leq s_{\rm f}$, except $s=0$.
For large $\tau$ one gets~\cite{ANlevelcrossing}
\BEA
\label{w7}
|a_{12}(\tf)|^2
=\frac{\pi\hbar\alpha_1^2}{4\tau\alpha_3^3},
\EEA
which for $\tau\to\infty$ is in accordance with 
the adiabatic theorem (\ref{abu}).  
From (\ref{suomi1}) we have 
$W-W_{\rm ad}=[\ep_{2}(\tf)-\ep_{1}(\tf)]\Theta_1$.
Using $\Theta_1=(p_1-p_2)|a_{12}(\tf)|^2>0$  we find,
because of the level crossing,  
$W-W_{\rm ad}=-|\ep_{2}(\tf)-\ep_{1}(\tf)|\,\Theta_1<0$, thus confirming
the violation of the minimum work principle. 
Eq.~(\ref{w7}) also shows that
the role of the proper internal characteristic time is 
$\hbar \alpha_1^2/\alpha_3^3$.

In the above two-level example one crossing was sufficient to violate
(\ref{2L}) with practically any non-adiabatic variation which produces
$\Theta_1>0$. For a many-level S having two of its levels crossed, any
non-adiabatic variation will not do anymore, as only one term in the
RHS of (\ref{suomi1}) will be negative. However, for quasi-adiabatic
variations -- and provided $\hat H(t)$ is analytic -- the 
transition probabilities between non-crossing levels are
exponentially small \cite{standardad,pechu}, while as we seen it has
power-law smallness for the two crossing levels.  For this situation
one can neglect in (\ref{suomi1}) the contribution from non-crossing
levels, and the problem is reduced to the two-level situation.  The
same concerns macroscopic systems which have at least two discrete
levels at the bottom of a continuous spectrum, whenever for low
temperatures these levels decouple from the rest of the spectrum.

\end{document}